\documentclass{aastex63}

\newcommand{\msun}{$\rm{M_\odot}$ }
\newcommand{\msunns}{$\rm{M_\odot}$}
\newcommand{\bvf}{Brunt-V\"ais\"al\"a }
\newcommand\mh{$\mathrm{M_H}$ }
\newcommand\mhns{$\mathrm{M_H}$}
\newcommand\mhe{$\mathrm{M_{He}}$ }
\newcommand\mhens{$\mathrm{M_{He}}$}
\newcommand\menv{$\mathrm{M_{env}}$ }
\newcommand\menvns{$\mathrm{M_{env}}$}

\shorttitle{Period spacings and Gaia data constrains for DAVs}
\shortauthors{Bischoff-Kim \& Bell}

\begin{document}

\title{Constraints from parallaxes and average period spacings in the asteroseismic study of 8 DAVs}
\correspondingauthor{Agn{\`e}s Bischoff-Kim}
\email{axk55@psu.edu}

\author[0000-0002-7487-9340]{Agn{\`e}s Bischoff-Kim}
\affil{Penn State Scranton \\
120 Ridge View Drive \\
Dunmore, PA 18412, USA}

\author[0000-0002-0656-032X]{Keaton J.\ Bell}
\affil{Department of Physics \\
Queens College \\
City University of New York \\
Flushing, NY-11367, USA}

\begin{abstract}
With space missions such as \emph{Kepler}, \emph{TESS}, and \emph{Gaia}, we have a wealth of data on pulsating white dwarfs that can be leveraged in white dwarf asteroseismology. We address the question of the proportion of white dwarfs with thin hydrogen layers versus those with thick hydrogen layers. We also provide a mass-radius relation for carbon-oxygen core, hydrogen atmosphere white dwarfs. Such a relationship can be used in conjunction with magnitudes and distance measurements to constrain the mass and effective temperature of the white dwarfs. We select eight hydrogen atmosphere, pulsating white dwarfs (DAVs), for their rich pulsation spectra. From such pulsation spectra, we can derive an asymptotic period spacing, which in turn allows us to determine the thickness of the hydrogen and helium envelope of the models, without having to perform period by period fitting. We find that the majority of the white dwarfs have thicker hydrogen layers and determine an upper limit of $M_r = 1 - 10^{-2.2}$ for the location of the base of the helium layer, in accordance with stellar evolution models. We confirm a finding from earlier studies that used a mass-radius relation and Gaia data to determine the effective temperatures of white dwarfs. The Gaia data systematically points to white dwarfs of lower effective temperature than indicated by the spectroscopy. Our results also support the hypothesis that white dwarfs with thicker hydrogen layers are more common than those with thinner layers. 
\end{abstract}

\section{Astrophysical Context}
\label{sec:introduction}

Recent space missions have contributed a vast amount of data on white dwarfs and pulsating white dwarfs. The \emph{Gaia} mission, launched in 2013, surveyed the entire sky and yielded distances to $\simeq$ 359000 white dwarfs to unprecedented precision \citep{GF21}. In parallel, planet finding missions have collected light curves of variable stars in partial or whole sky surveys. The \emph{Kepler} mission \citep{Borucki10}, operating between years 2009 and 2012, was perfect for white dwarf asteroseismology, providing observation baselines of fainter stars of up to a few years. The second \emph{Kepler} mission \citep[\emph{K2})][]{Howell14} focused along the ecliptic and continued to provide data on pulsating stars, if on shorter baselines. The \emph{K2} mission ended in 2018. The NASA mission following, the Transiting Exoplanet Survey Satellite \citep[\emph{TESS}][]{Ricker15}, focused on brighter stars but continued to provide a wealth of data on pulsating white dwarfs. To date we have spectroscopic, photometric, astrometric, and pulsation data for 500 hydrogen atmosphere white dwarfs \citep{Romero22,Guo23}.

White dwarfs are the end result of the evolution of lower mass stars \citep[ $\sim$ 98\% of all stars;][]{Althaus10b}. This means that they are common objects, and also that their interior structure holds the key to understanding the physical processes that take place in stellar evolution, such as nuclear fusion, core overshooting, mass-loss and diffusion \citep{Saumon22}. These processes set the shape of the internal chemical profiles of white dwarfs. White dwarf astereroseismology has long held the promise to unveil that internal structure \citep{Winget08,Fontaine08,Althaus10b,Corsico19}. Pulsations have been measured in white dwarfs since the discovery of the first DAV (hydrogen atmosphere pulsating white dwarf), HL Tau 76 \citep{Landolt68}. In the ensuing decades, work on the observational front to better measure and characterize the pulsations \citep[such as with the Whole Earth Telescope,][]{Nather90} progressed in parallel with work on the modeling side, with the aim of using the pulsations to infer the interior structure of white dwarfs \citep{Winget81}. Early efforts focused on determining the thicknesses of hydrogen and helium layers \citep{Bradley89,Kawaler90,Brassard92,Fontaine92}. Later, more sophisticated white dwarf models and fitting methods also focused on the carbon and oxygen abundance in the core \citep{Bradley98}. Much of that work was done by performing period by period fitting. The techniques on that front improved to the point where nearly perfect matches were found between the model periods and the observed period spectra \citep{Giammichele18}. There has been healthy scientific debates on the relative abundance of carbon and oxygen in the core of white dwarfs \citep[e.g.][]{Giammichele22,DeGeronimo19,Althaus10,Metcalfe02}.

When it comes to the helium-hydrogen envelope, and in particular the thickness of the hydrogen envelope, a picture has emerged where two types of DAs are formed: those with thick hydrogen envelopes and those with thin hydrogen envelopes. This hypothesis comes from the observation of a mismatch between the ratio of hydrogen to helium atmosphere white dwarfs below 15,000~K and that for lower temperature white dwarfs. In a systematic spectrosocopic survey, \citet{Tremblay08} found that 15\% of the DAs in the younger population would become DBs (helium atmosphere white dwarfs) once cooler. An explanation called ``spectral evolution" was proposed \citep{Fontaine87,Shibahashi07}, where for these white dwarfs, the hydrogen atmosphere was mixed by convection into the deeper, helium layers. Their atmospheres then become helium dominated, with trace amounts of hydrogen, if detectable at all. This is plausible, as DAs are found to pulsate between $\sim$ 13,000~K and 10,000~K, where conditions are ripe for the development of convection in the hydrogen layer. The most likely formation scenario for thin hydrogen atmosphere white dwarfs is mass-loss due to stellar winds after a late thermal pulse on the AGB, during a helium burning phase in the envelope \citep{Schoenberner79,Iben84,Althaus05b,Miller-Bertolami05}. In order for the hydrogen atmosphere of white dwarfs to be mixed out of existence, it must be thin to begin with. The existence of very thin hydrogen layers (\mh $< 10^{-14}$) have also been proposed \citep{Althaus20}. Such remnants would be the result of progenitors undergoing a Very Late Thermal Pulse episode. Thick ("canonical") hydrogen envelopes are the result of the standard evolution of a hydrogen rich remnant. 

Asteroseismology allows us to measure the thickness of hydrogen layers. All white dwarfs with hydrogen atmospheres are observed to pulsate as they evolve through the DAV instability strip \citep{Bergeron04,Castanheira07}, so studies of large samples of DAVs should allow us to verify the fact that $\sim$ 15\% of DAs have thin hydrogen layers. \citet{Miller-Bertolami17} define "thin" as \mh $<10^{-6}$, in a coordinate system where \mh $=M_H/M_* = 10^{-2}$ denotes a model where the pure hydrogen layer constitutes the outer 1\% of the model. The thickest hydrogen envelopes arising from stellar evolution calculations constitute only the outer 0.01\% of the models ($10^{-4}$) \citet[e.g.][]{Althaus10}. We use that as our higher limit for hydrogen layer masses when performing asteroseismic fits. For the purposes of this work, we shall use that definition of thick versus thin hydrogen layers and not single out very thin hydrogen layers. Several such studies have been conducted.  In a study of 44 DAVs, \citet{Romero12} found that 34 (77\%) had thicker hydrogen layers (\mh $>10^{-6}$). In a bigger study, the same group found that 43 of 74 DAVs (58 \%) had thicker hydrogen layers \citep{Romero22}. An independent study of 64 DAVs found the proportion of thicker hydrogen layers to be 38 \% \citep{Castanheira09}. \citet{Romero12} used white dwarf models derived from fully evolutionary models, with core chemical structures resulting from stellar evolution before the white dwarf stage and time dependent diffusion of elements on the white dwarf cooling track. \citet{Castanheira09} used White Dwarf Evolution Code \citep[WDEC;][]{Bischoff-Kim18a} static models with fixed carbon and oxygen profiles, set in that work to be a homogeneous 50/50 mix of carbon and oxygen.

In white dwarf asteroseismology, one sometimes runs into non-uniqueness of solutions, where two or more sets of chemical structure parameters fit the period spectrum equally well. This is the core-envelope symmetry problem \citep{Montgomery03} and it is the likely explanation for the wide discrepancy between the Romero et al. and the Castanheira et al. studies. They two studies used different carbon-oxygen core structures and so it is not surprising that they found different solutions when it came to the helium and hydrogen layer masses. A more recent study using the WDEC but with (still fixed) carbon-oxygen core profiles chosen to mimic those resulting from fully evolutionary models still found a majority of models with thinner hydrogen layers \citep[23 out of 29;][]{Hall23}.

The thickness of the hydrogen layer of DAVs is one topic of this work. We take a step back from the period-by-period asteroseismic fitting and instead focus on broader features of pulsation spectra, namely the average period spacing. This alone carries information about the parameters and structure of the white dwarfs, especially when it comes to the characteristics of the envelope. We do a systematic exploration of what we can learn from period spacings, and fold in further constraints from Gaia magnitudes and parallaxes. We focus on the study of 9 DAVs. One purpose of the study is to develop a method and share models that allow to place constraints on stellar parameters using the spectroscopy, the parallax, apparent magnitude, and pulsation spectrum. Parts of the method require the ability to measure an average period spacing and this is only possible for a handful of the known white dwarf pulsators, but other parts of the method are more broadly applicable.

In section \ref{sec:average_period_spacing} we discuss the theory of asymptotic period spacings, how we measure one for our chosen objects, and explore the effect of different parameters on the models' asymptotic period spacings. In section \ref{sec:mass-radius}, we discuss how one can level spectroscopy and parallaxes to help constrain the mass and effective temperature of white dwarfs. We introduce in that section a mass-radius relationship for carbon-oxygen core DA white dwarfs based on our models, a necessary ingredient of the method. We present and discuss our results in section \ref{sec:results} and conclude in section \ref{sec:conclusions}.

\section{Constraints from the average period spacing}
\label{sec:average_period_spacing}

The presence of a measurable regular period spacing in the pulsation spectrum allows us to constrain the mass and effective temperature of the white dwarfs, along a trend line in that plane. We discuss the theoretical foundation of this effect and examine the dependence of those trend lines on the different model parameters. In this study, we used the WDEC to produce our white dwarf models and calculate normal modes of oscillation. The code is detailed in \citet{Bischoff-Kim18a}. Two main updates were made. One is the use of a newer version of MESA for opacities and equations of state and the other is the ability to have oxygen further out in the envelope. The previously published version of the WDEC used opacities and equations of state from MESA version 8118 \citep{Paxton16}. The updated code interfaces with MESA version 22.11.1 \citep{Paxton22}. Between the two versions, there were updates done to the equations of state in MESA. They have a minute effect on the WDEC models. 

\subsection{Asymptotic theory}
\label{sec:asymptotic_theory}

The periods observed in DAVs are (non-radial) g-modes and range between $\sim$ 100 and 1500~s. Because of geometric cancellations \citep{Dziembowski77}, $\ell=1$ modes are expected to have the largest amplitudes, while modes with angular degrees greater than $\ell=2$ are unlikely to be observable. It is common practice to assume that all observed periods are either $\ell=1$ or $\ell=2$ modes, even though observations of higher angular degree modes have been reported \citep{Thompson08}. In the theory of non-radial stellar oscillations, if we assume long periods (and so low frequencies), we find that the periods of the g-modes are given by \citep{Unno89}

\begin{equation}
\label{eqn:eq1}
    P_{k,\ell}=\frac{k\pi}{[ \ell(\ell+1)]^{1/2}}\left[ \int_{r_1}^{r_2}\frac{N}{r}dr\right]^{-1},
\end{equation}

\noindent where N is the \bvf frequency integrated between the turning points of the modes, and $k$ is the radial overtone of the mode. For a given white dwarf model, the integral is a constant in the asymptotic limit of long periods (high radial order), and we see that the periods are proportional to the factor $1/[\ell(\ell+1)]^{1/2}$. This means that there is a constant period spacing associated with each given $\ell$. Furthermore, there is a factor of $\sqrt{3}$ between $\Delta P_{\ell=1}$ and $\Delta P_{\ell=2}$. Moving foward, we shall refer to the former as $\Delta P_{1}$ and the latter $\Delta P_{2}$ for short.

The integral is sensitive to how compressible the interior is and this in turn is determined by the mass and effective temperature of the white dwarf. The \bvf frequency can be written \citep{Brassard91}:
\begin{equation}
\label{eqn:eq2}
N^2=\frac{g^2 \rho}{P}\frac{\chi_T}{\chi_\rho}\left[\nabla_{ad}-\nabla-\frac{1}{\chi_T}\sum_{i=1}^{N-1}\chi_{X_i}\frac{d\ln X_i}{d\ln P} \right ]
\end{equation}

\noindent where
\[ \chi_\rho=\left(\frac{\partial \ln P}{\partial \ln \rho} \right )_{T,{\left \{ X_i \right \}}}, \]
\[ \chi_T=\left(\frac{\partial \ln P}{\partial \ln T} \right )_{\rho,{\left \{ X_i \right \}}}, \]
\[ \nabla_{ad}=\left(\frac{\partial \ln T}{\partial \ln P} \right )_{ad,{\left \{ X_i \right \}}},   \]
\[ \nabla=\frac{d \ln T}{d \ln P} , \]

\noindent and
\[ \chi_{X_i}=\left(\frac{\partial \ln P}{\partial \ln X_i} \right )_{\rho,T,{\left \{ X_j \ne i \right \}}}   .\]

More massive white dwarfs have a greater surface gravity, and therefore the \bvf frequency in their interior is larger. This leads to smaller period spacings (eqn. \ref{eqn:eq1}). White dwarfs with a lower effective temperature have interiors that are more degenerate. That is also true of more massive white dwarfs, but the effect of an increased gravity is more important. For more degenerate interiors, $\chi_T$ is smaller, as pressure and temperature are weakly coupled. this leads to a smaller \bvf frequency and greater period spacings. There is a degeneracy between the effects of temperature and mass, which translate to diagonal trends when graphing lines of constant period spacing in the mass-effective temperature plane (e.g. Fig \ref{fig:06_teff_mass_constraints1}). Composition also plays a role, as is made explicit in \citeauthor{Brassard91}'s formulation of the \bvf frequency (eqn. \ref{eqn:eq2}). As discussed in section \ref{sec:introduction}, the thickness of the hydrogen layer mass influences the period spacings. We expect the period spacings to be larger for thinner hydrogen envelopes \citep{Tassoul90,Uzundag23}. In this work, we also explore the effect of the thickness of the He/H envelope.

There is a known range of period spacing that we expect from our white dwarf models. For $\ell=1$ modes, the expected period spacing ranges from 38\~s to 46\~s, and for $\ell=2$ modes, 20\~s to 27 \~s. Those ranges are shaded in grey in Figs. \ref{fig:01_dp_kic} and \ref{fig:02_dp_g29}.This informs our work in section \ref{sec:measuring_dp}. The lower period modes are sensitive to the fact that white dwarfs are not homogeneous in density and chemical makeup and do not follow this regular spacing. This is called "mode trapping" \citep{Winget81}.

\subsection{Measuring an asymptotic period spacing}
\label{sec:measuring_dp}

Measuring the asymptotic period spacing in the pulsation spectrum of a white dwarf presents difficulties and very few objects have a conclusive spacing. First, a sufficient number of periods, especially at higher radial overtone ($k$) is required. These modes, however, are also the most sensitive to shifts in frequencies due to rotation. Rotation splits the frequency of $\ell=1$ modes into triplets and $\ell=2$ modes into quintuplets \citep{Unno89}, with the $m=0$ mode being in the middle. Occasionally, we observe the triplets and more rarely the quintuplets. This allows us to positively identify the angular degree $\ell$ of a mode, and also its central frequency. Most often however, we only observe one member of the multiplet, and we do not know that we are detecting the central peak. Most white dwarfs rotate with a period of order of a day \citep{Kawaler04,Corsico19}, leading the frequency splittings of $\sim 7 \mu$Hz for $\ell=1$ modes and $\sim 11 \mu$Hz for $\ell=2$ modes. In periods, this translates to a shift of $\sim$ 1 second for a 400~s mode, and $\sim$ 10 seconds for a 1000~s mode. While the longer period modes are expected to follow the regular, asymptotic period spacing, they are also more sensitive to rotation and our inability to identify the central peak in the multiplet. With a sufficient number of modes, however, some of that effect gets averaged out as some of the off center modes will have a higher period than the central mode and others will have a lower period.

We selected 9 DAVs with period spectra that contain a sufficient number of higher $k$ modes. They are listed in table \ref{tab:periodspacings}. For each object, we collected all the known periods. We list the aggregate period spectra and references in Appendix \ref{sec:appendixA}. We used as sources of object the catalogs published by \citet{Bognar16} and \citet{Hermes17}.

There are three statistical methods to determine a regular spacing in a period spectrum. Each essentially folds the period measurements in phase on a series of test spacings and quantifies how concentrated (suggesting alignment with an even spacing) versus spread out (not aligned with the test period) these measurement are in phase. The first method, the inverse variance (IV) test \citep{ODonoghue94}, quite simply reports the inverse of the variance the period measurements in phase when folded on each test period (larger values are more in line with an even spacing). 

The second method is the Kolmogovov-Smirnov (K-S) method \citep{Kawaler88}. It compares the cumulative distribution of periods folded on each test spacing with the uniform distribution. The statistic $\log{(Q)}$ describes how consistent these distributions are with each other; it is \textit{smaller} when the folded periods are less consistent with a uniform random distribution, suggesting a concentration of measurements folded on a tested spacing. The last method is to take a Fourier transform of the period spectrum \citep{Winget91}. This fits zero-centered sinusoids with different test periods to the period measurements, which are offset from the x-axis by a constant amount (similar to how a spectral window is computed). If the measurements follow an even spacing, the best-fit sine wave with that spacing as its period will fit with a higher amplitude (and power) as its positive peaks can reach up to the data points. Mathematically, this is equivalent to folding the measurements around a unit circle and computing the distance of the centroid from the center \citep{Eyer99}. Each method has its strengths and weaknesses, and statistical significance is difficult to assess. Since we know that periods should align with even spacings on average, it is common to compute all three tests and select the regular spacing that all three tests support in the expected range of period spacings for white dwarfs \citep[e.g.,][]{Corsico21}.

We show some sample results in Figs. \ref{fig:01_dp_kic} and \ref{fig:02_dp_g29}. The rest of the objects are featured in Appendix \ref{sec:appendixC}. The bottom panel shows an average of the results, where the K-S statistic is flipped (so high values support spacings), and the final average translated vertically so that the y axis begins at zero. The first example is KIC 220453225. For this star, it is easy to spot two clear peaks (or troughs), one at 40~s and another at 48~s (working at the 2 significant figures level). The 40~s periods spacing is consistent with a typical $\ell=1$ asymptotic period spacing, while 48~s is consistent with a harmonic of the corresponding $\ell=2$ spacing ($\sim 40/\sqrt(3)=23$~s). The second example, G~29-38, is not as clean cut. But if we limit ourselves to the range of possible $\ell=1$ spacings for an average mass DAV from our model grid ($\sim 38$ to $\sim 46$~s), two clear peaks emerge. The first one (41~s) is consistent with an $\ell=1$ for a model that has the effective temperature of G~29-38, while the 45~s peak is a candidate harmonic for an $\ell=2$ spacing. We discuss G~29-38 further below.

\begin{figure}[h!]
\epsscale{0.6}
\plotone{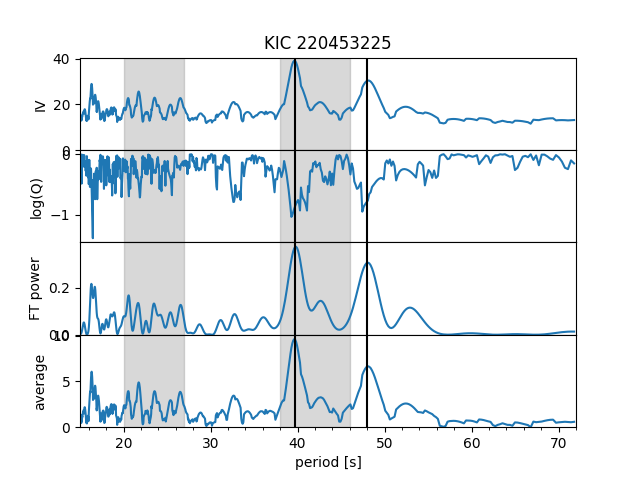}
\caption{Results of 3 statistical tests used in finding regular spacings in the period spectrum of KIC 220453225. The 4th panel is an average of the above curves (IV, -log(Q), and FT power). The grey regions mark the period ranges where we would expect a regular spacing corresponding to the asymptotic limit for $\ell=1$ modes at the higher period range and for $\ell=2$ modes at the lower period range. The peak around 48 s can be interpreted as the harmonic of a 24 s spacing, which would fall in the $\ell=2$ period spacing region.
\label{fig:01_dp_kic}
}
\end{figure}

\begin{figure}[h!]
\epsscale{0.6}
\plotone{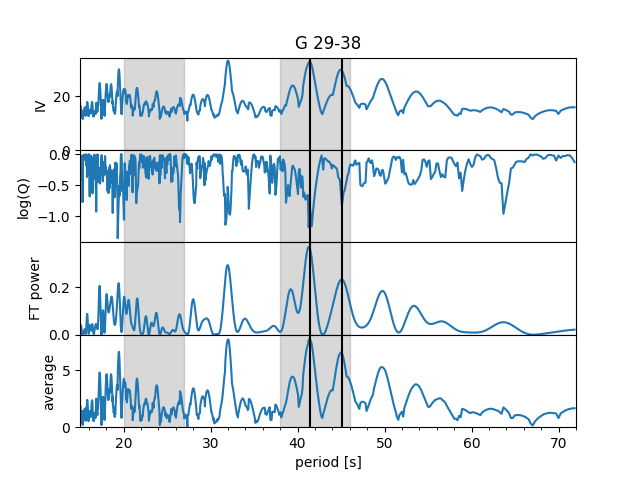}
\caption{The same as Fig. \ref{fig:01_dp_kic} for G 29-38. 
\label{fig:02_dp_g29}
}
\end{figure}

Some of our chosen targets were the object of previous studies. Aside for GD 1212, we recover the previously published results. For EC 14012-1446, \citet{Provencal12} found a period spacing of $\sim$ 41~s with a 99.99 \% confidence level of not being the result of a random process. \citet{Bell15} find a period spacing of 41.9 $\pm$ 0.2~s for KIC 4552982, by taking a Fourier transform of the period spectrum.

G~29-38 was the object of an asteroseismic study based on \emph{TESS} data \citep{Uzundag23}. In that work, 28 independent modes were found, along with multiplet structure that helped with the $\ell$ identification of the modes. Using sophisticated statistical methods, the authors determined an $\ell=1$ spacing for the star of $\sim$ 41~s and an $\ell=2$ spacing of 22~s. The K-S test confidence level for the $\ell=1$ spacing was nearly 100\%, while there was no significant peak corresponding to the $22$ ~ spacing. The $\ell=2$ spacing appeared in other tests used in the paper.
We can combine the results from that work with pulsation periods that were observed in G~29-38 in previous works to obtain a full list of periods and their $\ell$ identification. Many appear both in previous data and in TESS data. With the rich TESS pulsation spectrum, a good measurement of period spacings, and the multiplet structure observed in some of the modes, it is mostly a matter of filling the blanks in the $\ell=1$ and $\ell=2$ sequences. When calculating the slope of the the period versus $k$ graphs of the aggregate data, we find the same 41~s and 22~s period spacing for the $\ell=1$ and $\ell=2$ sequences respectively. The mode identifications that result from this work are given in table \ref{tab:period_lists}.

For GD 1212, \citet{Hermes14a} determined an $\ell=1$ period spacing of $41.5 \pm 2.5$ s. Our tests for GD 1212 are not conclusive. We observe a strong peak at $\sim 20$~s and nothing significant in the expected $\ell=1$ periods spacing region. It is possible that 20~s is a harmonic of 40~s. We adopted the published result of 41.5~s. 

KIC 201719578 has two complete triplets in its period spectrum, allowing to readily identify the $\ell=1$, $m=0$ mode periods \citep{Hermes17}. Combined with the results of the statistical tests described above, they allow us to form a likely $\ell=1$ sequence, with the remaining periods forming the $\ell=2$ sequence. As a result of the statistical tests, we find 3 peaks: 22~s, 45~s, and 24~s. From fitting a line to the likely $\ell=1$ sequence (period versus k), we find a period spacing of 45~s. When fitting a line through the remaining modes, we find a slope of 25~s. We adopt the latter as our $\ell=2$ period spacing. We list the period spacings of the white dwarfs we studied in table \ref{tab:periodspacings} and our mode identifications for this star in table \ref{tab:period_lists}.

R 808's period spectrum taken as a whole does not show any significant period spacing, but we can use 4 triplets to help us build an $\ell=1$ sequence. The rest of the modes show a spacing around 47~s, which \emph{can} be interpreted as twice an $\ell=2$ spacing. G38-29'spectrum does not show a consistent period spacing.

\begin{table}
  \begin{center}
  \caption{Period spacings for the stars considered in this study. $\ell=2$ spacings are not always found. G 38-29 does not show either an $\ell=1$ or $\ell=2$ period spacing.}
     \label{tab:periodspacings}
 {\scriptsize
  \begin{tabular}{|l|c|c|}
  \hline 
    Object          & $\ell=1$ period spacing   & $\ell=2$ period spacing   \\
    \hline
    EC 14012-1446   & 42~s                      &                           \\
    G 38-29         & None found                &                           \\ 
    R 808           & 40.5~s                    & 23.5~s                    \\
    KIC 4552982     & 41.9~s                    &                           \\
    G 29-38         & 41~s                      & 22~s                      \\
    GD 1212         & 42~s                      &                           \\
    KIC 210397465   & 52~s                      &                           \\
    KIC 220453225   & 40~s                      & 24~s                      \\
    KIC 201719578   & 45~s                      & 25~s                      \\
    \hline
 \end{tabular}
 }
 \end{center}
\vspace{1mm}
\end{table}

The period spacings extracted from the observed period spectra are then compared with the period spacings of the models. For the models, calculating an average period spacing is easier, as we have the luxury of complete sequences of normal modes. To obtain the average period spacing, we simply calculate the best fit line that goes through the points $(k,P_k)$, discarding the modes with $k<10$. The slope of the line gives us the period spacing. The code checks that there are a sufficient numbers of points and calculates a coefficient of linear correlation that helps verify that we have captured the modes that reside reasonably near the asymptotic limit.

\subsection{Dependence of the period spacing on model structure}
\label{sec:dp_dep_on_struc}

\begin{figure}[h!]
\epsscale{0.6}
\plotone{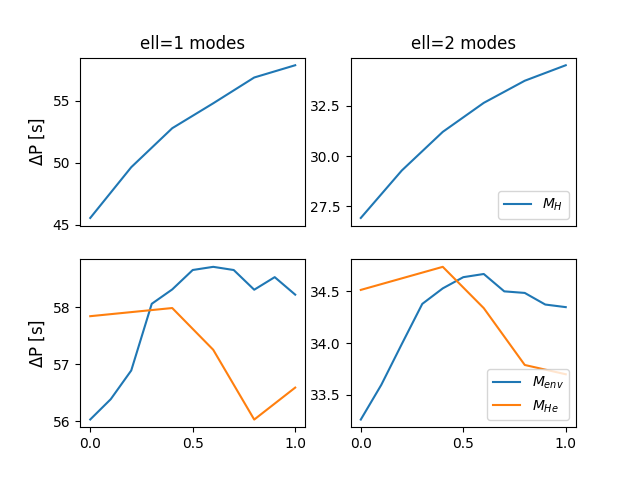}
\caption{Dependence of $\ell=1$ and $\ell=2$ period spacings on the thickness of the hydrogen layer (\mhns), helium and hydrogen envelope (\menvns), and pure helium layer (\mhens). The horizontal axis is normalized and spans the entire allowed range of each parameter. The axes run from thicker envelopes to thinner. For \menvns, the range is [$10^{-1.5},10^{-5.5}$], for \mhe [$10^{-2},10^{-7}$], and for \mh [$10^{-4},10^{-9}$]}.
\label{fig:03_dpdep1}

\end{figure}

\begin{figure}[h!]
\epsscale{0.6}
\plotone{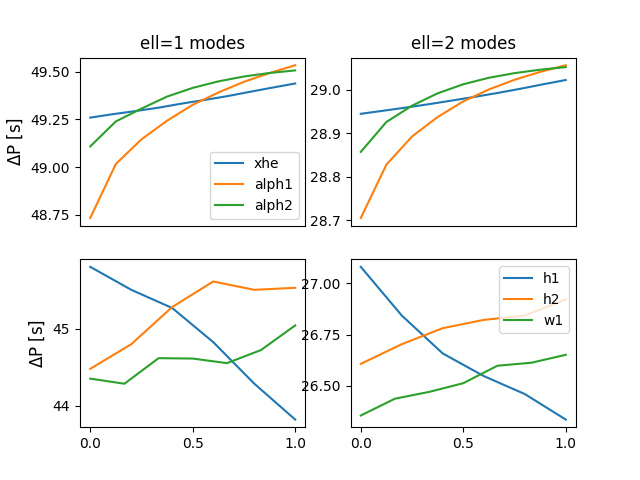}
\caption{Dependence of $\ell=1$ and $\ell=2$ period spacings on more envelope parameters (xhe, alph1, and alph2) as well as parameters that have to do with the oxygen abundance profile (h1, h2, w1). For xhe, the range is [0.1,1.0], for alph1 and alph2 [4,20], for h1 [0.5,1.0], for h2 [0.1,0.6], and for w1 [0.1,0.4]. xhe, h1, and h2 are relative chemical abundances, alph1 and alph2 diffusion exponents, and w1 a mass coordinate. See text for details. 
}
\label{fig:04_dpdep2}

\end{figure}

\begin{figure}[h!]
\epsscale{0.5}
\plotone{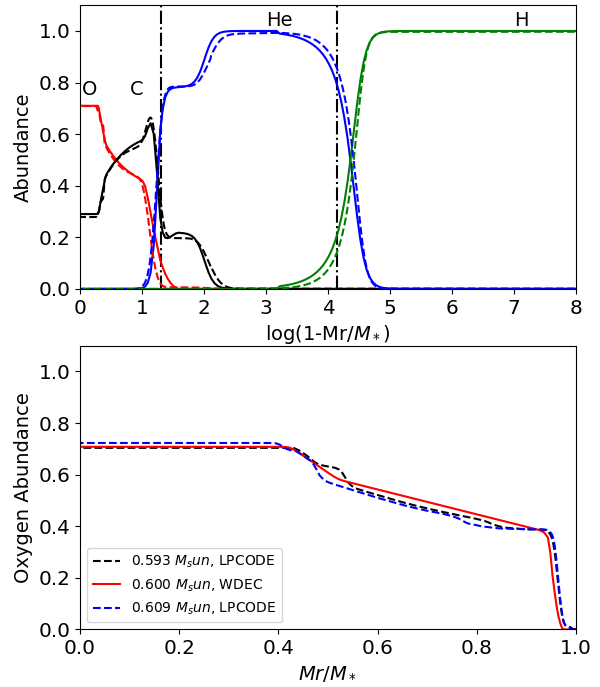}
\caption{Top panel: the chemical profiles of a 0.525 \msun white dwarf model computed with the LPCODE (dashed curves) and the WDEC (solid curves). The vertical, dash dotted lines mark the location of the base of the helium-hydrogen envelope (\menv $=10^{-1.31}$) and the location of the base of the hydrogen layer (a.k.a. hydrogen layer mass, \mh $=10^{-4.14}$). Bottom panel: Oxygen abundance profiles for 3 different models close to 0.6 \msunns.
\label{fig:chemprofiles}
}
\end{figure}

In section \ref{sec:asymptotic_theory} we oversimplified when we stated that mean period spacing depends only on mass and effective temperature. Outer envelope structure also matters. One would expect a thicker hydrogen layer to lead to a star that is overall more compressible than one with with a thin envelope. To verify that hypothesis, we studied the dependence of the period spacings on the parameters that describe the chemical structure of WDEC models. We show the results of that work in figures \ref{fig:03_dpdep1} and \ref{fig:04_dpdep2}, for both the $\ell=1$ and $\ell=2$ period spacings. We normalized the horizontal axes such that 0 corresponds to the minimum value of the parameters possible and 1 corresponds to the maximum possible value. In essence, each panel shows the variation in period spacing for the entire span of the given parameter. Normalizing allows us to put more than one parameter on one plot. We find that the parameters that set the oxygen profile (i.e. the chemical structure in the core, see Fig. \ref{fig:chemprofiles}), have an effect on the period spacings at the 2 to 3 percent level. Parameters that determine the smoothness of the transition between the C/O/He mix and pure helium layer (alph1 and alph2), as well as the parameter that sets the helium abundance in the homogeneous C/He mixed region (xhe) matter at the 1\% level. The parameter that sets the location of the transitions from the C/O core to the He/H envelope (\menvns) and the one that set the location of the transition to pure helium (\mhens) have a greater but still small effect ($\sim$ 5\%). All of these pale in comparison to the effect the thickness of the hydrogen layer (\mhns) has (20\%).

In light of those results, we proceeded by fixing the core structures. We adopted the chemical profiles of \citet{Althaus10}, generated using the LPCODE. These indeed depend on the mass of the white dwarf. Given the negligible effect the interior profiles have on the period spacings, it is fine to have varying interior profiles and beneficial to have them fixed to what detailed stellar evolution calculations predict. As a sample, we show the chemical profiles for a 0.525 \msun model, both from the LPCODE and from WDEC (top panel of Fig. \ref{fig:chemprofiles}). For masses not on the \citeauthor{Althaus10} mass grid, we interpolated the 7 WDEC parameters that set the oxygen profiles. The bottom panel of Fig. \ref{fig:chemprofiles} is an illustration of the result of such interpolation, done for a 0.6 \msun white dwarf.

\section{Constraints from spectroscopy and parallaxes}
\label{sec:mass-radius}

\citet{Bischoff-Kim23b} describe a method, based on basic principles, that allows one to use G-magnitudes and distances from the Gaia mission, as well as surface gravities from spectroscopy to obtain constraints in the mass-effective temperature plane. Such constraints appear in the form of diagonal lines that run nearly perpendicular to the period spacing lines (e.g. Fig. \ref{fig:06_teff_mass_constraints1}). A key component required in the calculation of such lines is a mass-radius relationship. \citet{Bischoff-Kim23b} provide such relationship for helium core white dwarfs. Here we want to construct one from WDEC models for carbon-oxygen core white dwarfs. We begin by determining what parameters influence the mass-radius relationship the most. Most of the parameters have no effect on the mass-radius relationship. This is particularly true of the parameters that describe the oxygen profile. White dwarfs are very compact and details of core composition have a negligible effect on their size. This allows us to use same approach to oxygen profiles as we did in section \ref{sec:dp_dep_on_struc} for asymptotic period spacings calculations. 

\begin{figure}[h!]
\epsscale{0.8}
\plotone{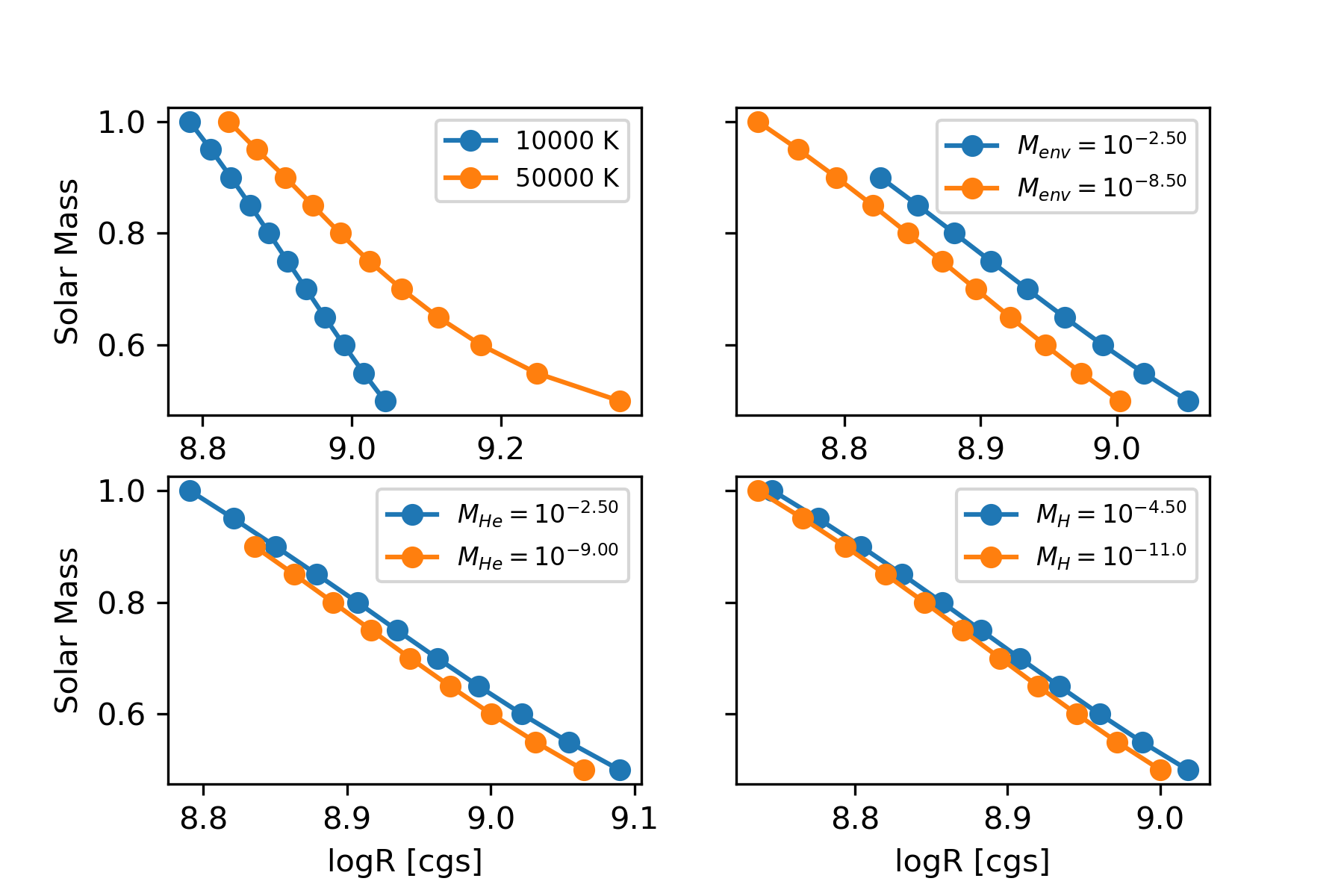}
\caption{Dependence of the mass-radius relation on the parameters that matter: effective temperature and outer envelope parameters. 
\label{fig:05_mass_radius_pardep}
}
\end{figure}

In Fig. \ref{fig:05_mass_radius_pardep} we show the effect of the 4 parameters that do affect the mass-radius relationship. Not surprisingly, temperature has the greatest effect, followed not by the thickness of the outermost layer (hydrogen), but rather by the thickness of the He/H envelope. We therefore construct a mass-radius relation that is a function of effective temperature and envelope mass \menvns. The relation is detailed in Appendix \ref{sec:appendixD}.

\section{Results and Discussion}
\label{sec:results}

We are ready to bring the constraints from period spacings and from distances together with spectroscopy to study our chosen white dwarfs. Figs.\ \ref{fig:06_teff_mass_constraints1}-\ref{fig:08_teff_mass_constraints3} show how these measurements constrain the masses and effective temperatures of these stars for different assumed envelope and hydrogen layer masses. Lines of constant period spacings matching the measured spacings were computed based on grids of models; these traverse the parameter space from high mass at low temperature, to low mass at high temperature. Lines of constant absolute magnitude, in agreement with the \textit{Gaia} measurements, follow an opposing trend in parameter space. We varied both the hydrogen layer and the He/H envelope masses of the models. In section \ref{sec:dp_dep_on_struc}, we determined that the period spacing was most sensitive to the hydrogen layer mass, and in section \ref{sec:mass-radius} we saw that the mass-radius relation, and therefore the constraints from Gaia magnitudes, parallaxes, and spectroscopy were most sensitive to the He/H envelope mass; therefore, we consider the effect of these two parameters (\menv and \mhns) in our study. Different combinations of \menv and \mhns\ are indicated with different line styles and colors in the figures. Constraints from period spacings and astrometry both agree with the models where lines of the same color and style intersect. Finally, the constraints from spectroscopy are shown as black points, surrounded with one-sigma error boxes. We list the effective temperatures and mass from spectroscopy in tables \ref{tab:spectroscopy1} and \ref{tab:spectroscopy2} (Appendix \ref{sec:appendixB}), along with the relevant references. When there was more than one determination for each star, we computed an average and a standard deviation for the mass and effective temperatures.

We begin by noting the two main results 1) The constraints from period spacings intersect with corresponding Gaia lines closest to the spectroscopic measurements for models with thicker hydrogen layer masses. They pass through the spectroscopic boxes for 7 of the 9 stars. 2) We observe a systematic offset between the spectroscopic mass and effective temperature of the white dwarfs and where we would expect them to fall according to the Gaia data. There are two exceptions: G 29-38 and GD 1212. Below, we discuss each of the 9 stars in turn.

\subsection{EC 14012-1446}
\label{sec:ec14012}

Based on the period spacing lines that cross the spectroscopic box, we find a range of possible envelope mass, from $10^{-1.8}$ to $10^{-4.0}$ (focusing on the style of the lines), and hydrogen layer masses from $10^{-4.0}$ to $10^{-5.5}$ line colors. The Gaia lines corresponding to the range of envelope masses quoted above all run to the left (lower effective temperature) of the spectroscopic box. The Gaia lines, being dependent on the mass-radius relation, are sensitive to the envelope mass, but to a much lesser extent on the hydrogen mass (and so that was not varied). For each layer mass, however we had to select a hydrogen layer mass. We chose the thickest allowed hydrogen layer each time.

\subsection{G29-38}
\label{sec:g2938}

Like EC 14012-1446, the range of envelope mass goes from $10^{-1.8}$ to $10^{-4.0}$, and hydrogen layer masses go from $10^{-4.0}$ to $10^{-5.5}$. This is in agreement with the full asteroseismic fitting performed by \citet{Uzundag23}. Their preferred solution, based on combining all known periods for G29-38, points to a hydrogen layer mass of $7.58 \times 10^{-5}$ and an envelope mass of $1.74 \times 10^{-2}$. Their best fit model has a mass of 0.632 \msunns. G29-38 is known to have a debris disk around it. The disk is evidenced both by the presence of metals in G 29-38's spectrum (it is a DAZ), and by an infrared excess \citep{Zuckerman87}. Subsequently, it was found that debris disks were not uncommon around white dwarfs \citep{Kilic06}. An optically thick debris disk would dim the white dwarf, and we would infer that it is smaller than it really is, and therefore more massive. This effect would move the Gaia lines to higher mass, toward the upper left corner of the graph in Fig. \ref{fig:06_teff_mass_constraints1}. For G29-38, the Gaia lines intersect the spectroscopic box, in contrast with what we observe for all of the other objects in this study. If we were seeing the effect of the debris disk around G29-38, we should see the opposite (the discrepancy should be even greater).

One striking feature of Fig. \ref{fig:06_teff_mass_constraints1} is the discrepancy between the picture for the $\ell=1$ versus the $\ell=2$ modes. This is a good illustration of the fact that in order to determine a meaningful asymptotic spacing from an observed period spectrum, not only are longer period modes necessary, but a sufficient number of them as well. In the case of GD 29-38, we have 15 $\ell=1$ modes at our disposal, but only 3 $\ell=2$ modes. While we recover the period spacings and mode identifications of \citet{Uzundag23} in our analysis, we do not obtain a useful measure of the asymptotic period spacing for the $\ell=2$ modes. We note that obtaining an $\ell=2$ period spacing is still worthwhile to do, as it helps in identifying modes when performing an asteroseismic fit, as \citet{Uzundag23} did.

\subsection{GD 1212}
\label{sec:gd1212}

\begin{figure}[h!]
\epsscale{1.0}
\plotone{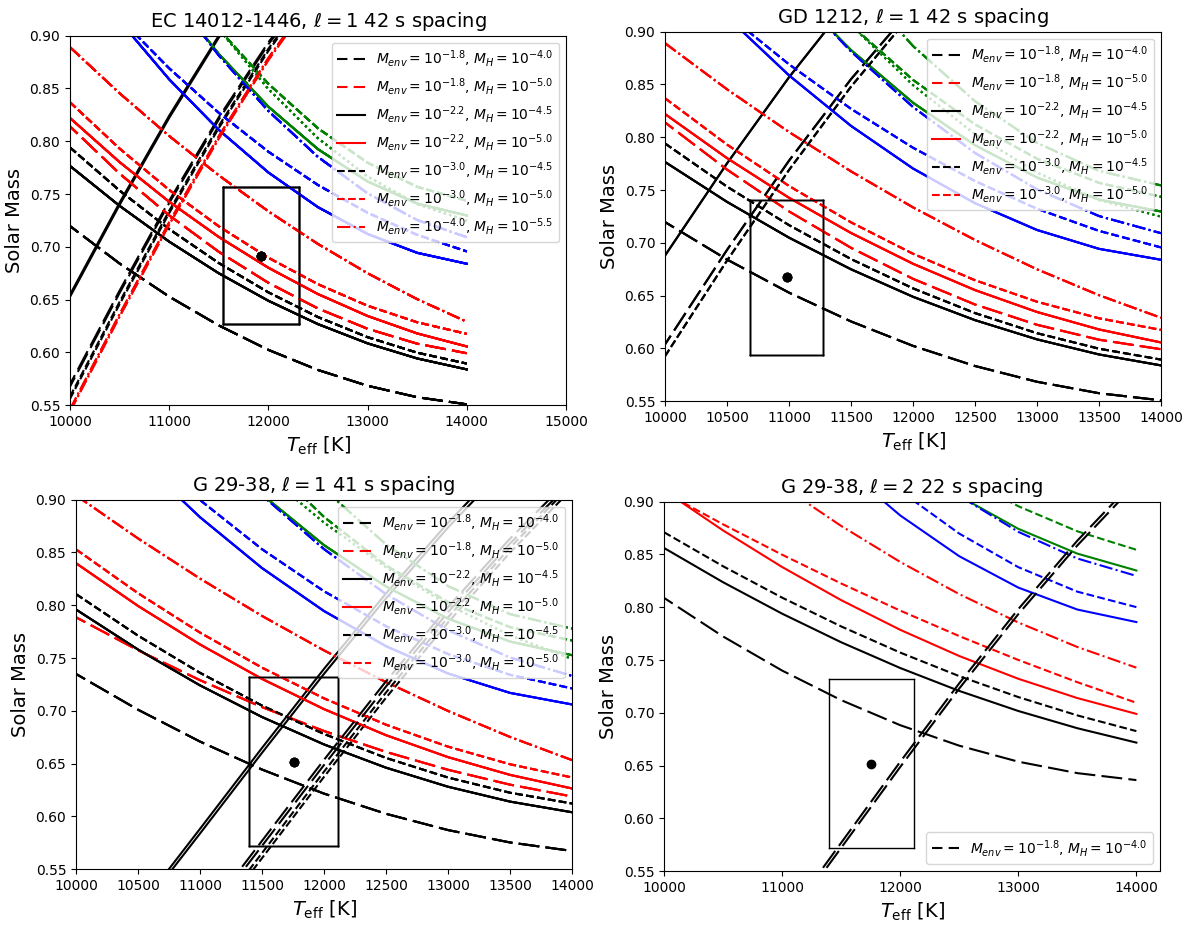}
\caption{Constant period spacing lines (negative slopes) and constraints from spectroscopy combined with Gaia magnitudes and distances (positive slopes) for EC 14012-1446, GD 1212, and G 29-38. Even though EC 14012-1446 and GD 1212 exhibit the same $\ell=1$ period spacings (and so the period spacing lines are identical), we chose to put them on separate plots to reduce clutter. In each panel, the solid circle marks the location of the average effective temperature and mass with a one standard deviation rectangle around it. The constant period lines begin at lower stellar masses with the thickest hydrogen layers (\mh $=10^{-4.0}$ and $10^{-4.5}$) and progress to thinner and thinner envelope masses. The different line styles correspond to different hydrogen layer masses. We includes lines of constant period spacing for thinner hydrogen layers \mh = $10^{-7}$ (blue) and $10^{-9}$ (green) for completeness, but none of the objects we include in this study have measured properties consistent with such thin hydrogen layers. They are not listed in the legends but follow the same line style scheme as the lines shown in the legends.
}
\label{fig:06_teff_mass_constraints1}

\end{figure}

\begin{figure}[h!]
\epsscale{1.0}
\plotone{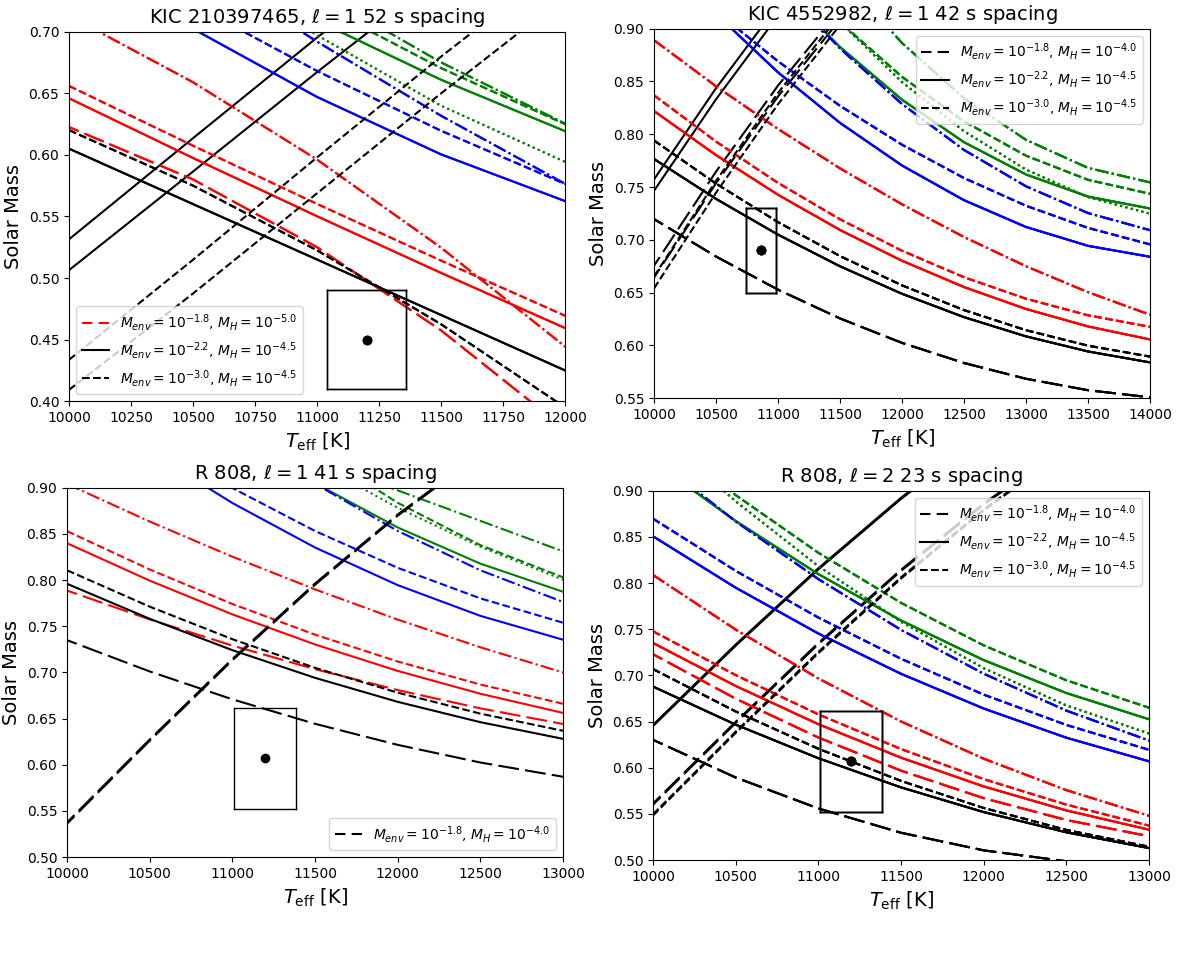}
\caption{Same as Fig. \ref{fig:06_teff_mass_constraints1} for KIC 210397465}, KIC 4552982, and R 808.
\label{fig:06_teff_mass_constraints2}
\end{figure}

\begin{figure}[h!]
\epsscale{1.0}
\plotone{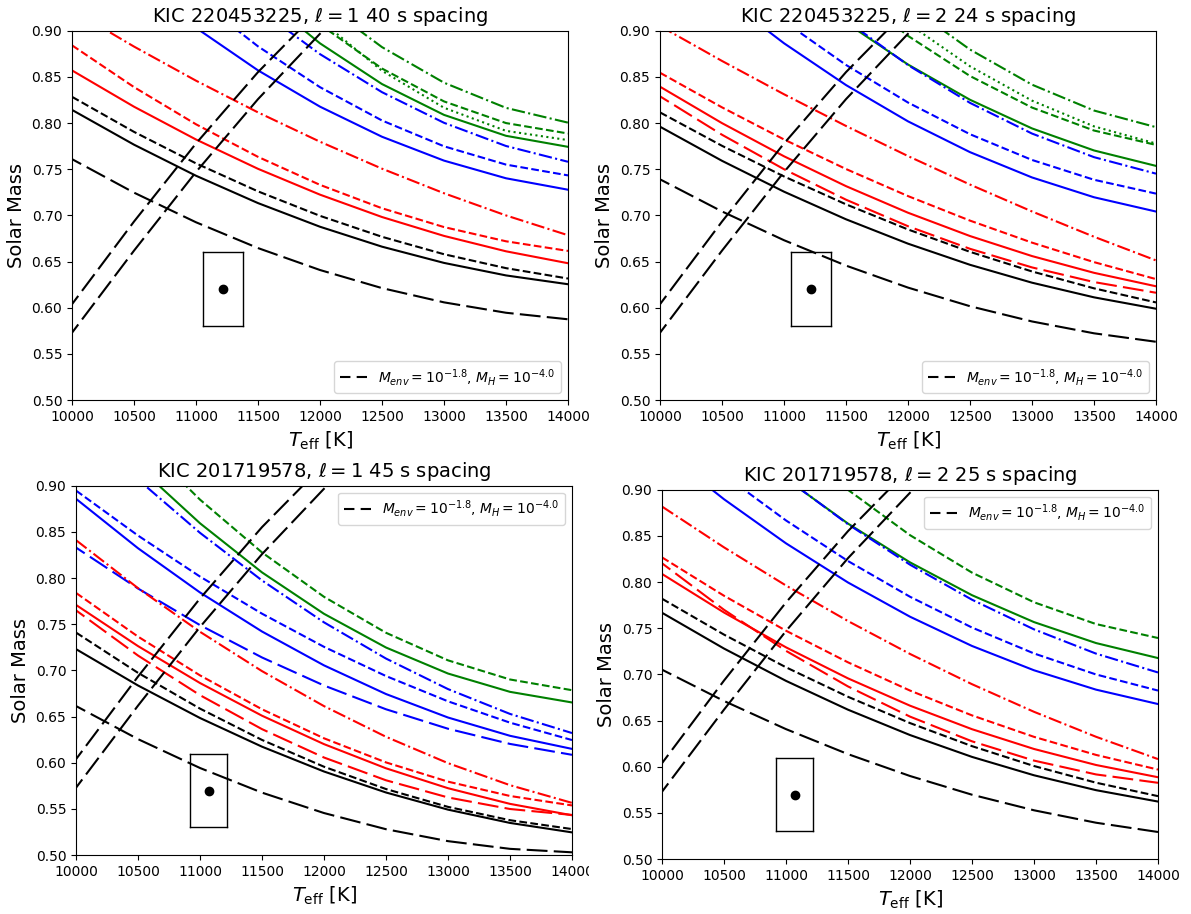}
\caption{Same as Fig. \ref{fig:06_teff_mass_constraints1} for KIC 220453225 and KIC 201719578.
\label{fig:08_teff_mass_constraints3}
}
\end{figure}

We again have a similar picture for GD 1212. Taking the period spacing and Gaia lines together, we find an envelope mass between $10^{-1.8}$ and $10^{-2.2}$, and a hydrogen layer mass between $10^{-4.0}$ and $10^{-4.5}$ to be consistent with 1-sigma spectroscopic error boxes. GD 1212 was the object of asteroseismic fits \citep{Romero17} using the LPCODE \citep{Corsico06} and the WDEC. Both studies found a higher mass for the star (0.815 and 0.877 \msun respectively). Both found thick (\mh $> 10^{-6}$) hydrogen layers ($3.5 - 7.4 \times 10^{-6}$ and $10^{-4}$ respectively). The relevant curve in Fig. \ref{fig:06_teff_mass_constraints1} is the black, long dash curve corresponding to an envelope mass of $10^{-1.8}$ and hydrogen layer mass of $10^{-4}$. Furthermore, the best fit WDEC model in \citet{Romero17} has an effective temperature of $11100$~K. In Fig. \ref{fig:06_teff_mass_constraints2}, the corresponding mass coordinate on the \menv $=10^{-1.8}$ curve is $\sim 0.65$ \msunns. This is in better agreement with the spectroscopy. The fact that the curve corresponds to a 42~s period spacing instead of 41.5~s does not explain the discrepancy. There is only a small shift between the 41~s and 42~s constant period lines (e.g. left two panels in Fig. \ref{fig:06_teff_mass_constraints1}). It is more likely the fact that the WDEC models used in \citet{Romero17} have a different envelope structure, which does not have a region of homogeneous He-C mix \citep{Castanheira08}.

\subsection{KIC 4552982}
\label{sec:kic4552982}

KIC 4552982 was the object of a nearly continuous, one and a half year-long observing run by the \emph{Kepler} satellite. It was the first pulsating white dwarf that was directly observed to have semi-periodic, pulsationally driven outbursts \citep{Bell15}. Its pulsation spectrum alone is of note, by its richness and regularity. From the 20 modes observed, \citet{Bell15} was able to directly measure a period spacing by taking a Fourier transform of the Fourier transform (in period space). In the same work, the period spacing was used to arrive at the conclusion that 1) the star was over 0.6 \msun in mass, and that the most likely hydrogen layer mass was $10^{-4.5}$. We reaffirm both findings.

\subsection{R 808}
\label{g38-29 and r808}

For R 808, the $\ell=1$ period spacing lines point to thicker envelope and hydrogen layers ($10^{-1.8} \leq$ \menv and $10^{-4.0} \leq$ \mh ), while the $\ell=2$ period spacing lines allow for thinner envelope masses ($10^{-3.0} \leq$ \menv $\leq 10^{-1.8}$, $10^{-4.5} \leq$ \mh $\leq 10^{-4.0}$).

\subsection{KIC 220453225, KIC 201719578, and KIC 210397465}

For KIC 220453225, the 40~s $\ell=1$ period spacing lines miss the 1-sigma spectroscopic box, but the line corresponding to the thickest envelope and hydrogen mass (\menv $= 10^{-1.8}$ and \mh  $= 10^{-4.0}$) passes within 2 sigma of the spectroscopic measurement. For the 24~s $\ell=2$ period spacing, that line grazes the corner of the 1 sigma spectroscopic box.

For KIC 201719578, we again have a preference for thicker envelope masses. We also observe a discrepancy between the $\ell=1$ and $\ell=2$ period spacing. Unlike G 29-38, we do have a rich spectrum of both possible $\ell=1$ and $\ell=2$ modes. Those spacings appear as peaks in the statistical tests of section \ref{sec:average_period_spacing} as well as in more detailed analysis. The discrepancy does not, however, change the conclusion that thicker envelopes are preferred.

According to the spectroscopy, KIC 210397465 is a lower mass white dwarf ($0.45 \pm 0.04 $ \msun), and we do find a longer period spacing consistent with a lower mass DAV. Such models are at the lower mass end of our grid and the period spacing lines are more challenging to calculate. However, they are consistent with thicker envelope and hydrogen layer masses.

\bigskip
For 7 out of the 8 DAVs considered here, we observe that constraints from Gaia magnitudes and parallaxes are systematically cooler and more massive (smaller radius) than values inferred from spectroscopy. This can be interpreted as a systematic error in apparent magnitude, with objects measured to be systematically dimmer than they really are (by $\sim$ 0.4 in magnitude). This would make them appear more massive, shifting the Gaia lines upward in the figures. Alternatively, it could be an error in the spectroscopy, either with surface gravities underestimated, or effective temperatures overestimated. \citet{Tremblay19} have considered this question, by studying a sample of 7039 DAs (and 721 DBs). For the DAs, \citeauthor{Tremblay19} conclude that there is a systematic difference in effective temperature between those determined from Gaia magnitudes and parallaxes and that from spectroscopy. They observe that the Gaia effective temperatures are systematically smaller than the spectroscopic values. This effect is seen both for the spectroscopy of the DA sample of \citet{Gianninas11} corrected for 3D effects \citep{Tremblay13} and to a greater extent for the SDSS DR7 sample \citep{Abazajian09}. They do not quantify the magnitude of the shift, but quantitatively, we see the same effect in the present work. The authors conclude that the more likely explanation for this discrepancy is residual issues with the spectroscopic determinations of effective temperature. They do not see a systematic difference in surface gravities. Given the sharply peaked mass distribution of the sample, if surface gravities from spectroscopy are wrong as well, the physical range is too small for the effect to be seen.

\section{Summary and Conclusion}
\label{sec:conclusions}

We set out to address the question of the fraction of thick versus thin hydrogen layers in DA white dwarfs, with an expectation that the majority would have thick hydrogen layers. From a study purely based on the asymptotic period spacing, effective temperatures (from spectroscopy), and Gaia data (absolute magnitude and parallaxes) of selected DAVs, we find that overall, models with thicker envelopes are closer to constraints from spectroscopy than models with thinner envelopes, suggesting that WDs with thick hydrogen layers are more common. This is in accordance with the results of \citet{Romero12} and \citet{Romero22}, who found that 77\% and 58 \% respectively of the DAVs under study had thick hydrogen layers. We also recover the thicker helium/hydrogen envelopes expected from stellar evolution, with an upper limit of $10^{-2.2}$ for the location of the base of the helium layer in all of the DAVs considered in this study. This is to be taken with the caveat that we also confirm a result of \citep{Tremblay19}, who suspect that the effective temperatures from spectroscopy are too high to be consistent with the distances measured by Gaia. However, shifting the spectroscopic boxes to lower effective temperatures only solidifies the conclusion that for the 9 objects considered in this study, thicker helium/hydrogen envelope masses match the constraints better.

If there are enough long period modes to measure an asymptotic period spacing, such spacing is a good starting point or any period-by-period fitting as it can aid in the $\ell$ identification of the modes. An illustration of this is the study of the DBV (helium atmosphere pulsating white dwarf) TIC 257459955 \citep{Bell19}. In that study, the best fits had an effective temperature of $\sim$ 30,000~K, but ultimately the authors selected as a solution a best fit model that had an effective temperature of $\sim$ 25,000~K, in agreement with the spectroscopy for the object. It is useful to explore the wider parameter space and report all best fits, even if not all match the spectroscopy or other observables we have for the object. It is also good when one can bring to bear as many external constraints as possible on asteroseismic fits. The work done here on DAVs that have enough periods in their pulsation spectrum to allow us to determine asymptotic period spacings can inform more detailed asteroseismic fittings of the objects. 

The WDEC is open source and may be obtained from GitHub \citep[][Codebase: \\ \url{https://github.com/kim554/wdec}]{Bischoff-Kim18b}. Further documentation, including a user manual, may be found in the GitHub repository. In this study, we also used the utility at \url{https://github.com/keatonb/meanperiodspacing} to calculate period spacings.

\software{WDEC \citep{Bischoff-Kim18b}}

\acknowledgments

The authors thank the referee for thorough, constructive and timely feedback. 

\bibliographystyle{aasjournal}
\bibliography{index}

\begin{appendix}

\section{Pulsation data}
\label{sec:appendixA}

We provide the period spectra for each object considered in this study in table \ref{tab:period_lists}. To calculate period spacings, we did not include periods under 600~s, with the exception of G 29-38, where we included periods as low as 400~s (to compare with work done by \citet{Uzundag23}). Considering the number of modes present in this period spectrum, the mode trapping that occurs for the lower $k$ modes has a negligible effect on the asymptotic period spacing. For most objects, we did not perform a detailed analysis that would allow us to identify the modes.  The sources cited sometimes offer a mode identification. A knowledge of the angular degree $\ell$ of the modes can allow to refine the values of the period spacings but is not absolutely necessary for the purposes of this work.

 \begin{table}
  \begin{center}
  \caption{Period spectra used in this work. When appropriate, mode identification is indicated in parentheses ($l$) or ($l$,$k$,$m$). Periods with an asterisk next to them are an average of members of a multiplet. We list all of the periods available for each object, including lower $k$ modes not used in the determination of period spacings. When periods appeared in more than one observing run, we adopted the value from the latest run.
     \label{tab:period_lists}
}
 {\scriptsize
  \begin{tabular}{lcl}
  \hline 
Object              & Period list                                                                                           & Reference             \\
\hline
EC 14012-1446       & See table 4 of \citet{Provencal12}                                                                    & \citet{Provencal12}   \\
                    &                                                                                                       &                       \\
G 38-29             & 707.601$^*$, 840.39, 899.971, 922.657, 945.448, 962.835$^*$, 984.93$^*$, 1002.16, 1016.15, 1085.99$^*$    & \citet{Thompson09b}   \\
                    & not used: 413.307, 432.354, 546.96                                                                    &                       \\
                    &                                                                                                       &                       \\
R 808               & 630.7035$^*$ (1), 745.12 (1), 796.253 ((1), 842.707 (1), 860.227, 876.8125$^*$ (1), 898.707, 908.422,  & \citet{Thompson09b}   \\
                    &   914.0$^*$ (1), 922.504, 952.398, 960.527 (1), 1011.39, 1041.06$^*$ (1) , 1066.73 (1), 1091.09, & \\
                    & 1143.96, 2459.1 & \\                                                                                            &   \\
                    & not used: 404.457, 511.266                                                                            &                       \\
                    &                                                                                                       &                       \\
KIC 4552982         & See table 2 of \citet{Bell15}                                                                         & \citet{Bell15}        \\
                    &                                                                                                       &                       \\
G 29-38             & 400.452 (1,13,0), 449.689 (1,14,0), 488.780 (1,15,0), 571.219 (1,17,0), 610.668 (1,18,0),             & \citet{Uzundag23}     \\
                    & 655.055 (1,19,0), 713.477 (1,21,0), 846.270 (1,24,0), 899.326 (1,25,0), 939.553 (1,26,0),             &                       \\
                    & 1157.358 (1,31,0), 1192.051 (1,32,0), 1351.244 (1,36,0)                                               &                       \\
                    & not used: 495.879 (2,25,-2)                                                                           &                       \\
                    &                                                                                                       &                       \\
                    & 770.8 (1,22,0), 809.4(1,23,0), 1239.9 (2,58,0)                                                        & \citet{Castanheira09} \\
                    & not used: 218.7, 283.9 (1,10,0), 363.5 (1,12,0), 894.0 (1,25,-1), 1150.5 (1,31,-1), 1185.6 (1,32,-1)  &                       \\
                    &                                                                                                       &                       \\
                    & 915 (2,44,0), 1147.0 (2,54,0)                                                                         & \citet{Kleinman98}    \\
                    & not used: 110.0 (1,6,0), 177.0, 237.0 (1,9,0), 355.0 (1,12,0), 552 (2,26,2), 678.0 (2,33,1)           &                       \\
                    &                                                                                                       &                       \\
GD 1212             & 828.19, 849.87$^*$, 871.06, 956.87, 987.00, 1008.07, 1025.31, 1048.19, 1063.08, 1086.12, 1098.36,     & \citet{Hermes14a}      \\
                    & 1125.37, 1147.12, 1166.67, 1180.4, 1190.53, 1220.75                                                   &                       \\
                    & not used: 369.83, 371.05                                                                              &                       \\
                    &                                                                                                       &                       \\
KIC 210397465       & 600.73$^*$, 667.014, 710.553, 758.46, 977.03, 1019.00, 1072.39, 1220.306, 1278.1, 1382.26             & \citet{Hermes17}      \\
                    &                                                                                                       &                       \\
KIC 220453225       & 670.886, 831.942, 875.655, 918.2, 960.794, 997.60, 1039.242, 1072.6, 1107.1, 1156.71,                 & \citet{Hermes17}      \\
                    & 1204.1, 1257.00, 1304.50, 1353.86, 1391.46                                                            &                       \\
                    & not used: 311.4643                                                                                    &                       \\
                    &                                                                                                       &                       \\
KIC 201719578       & 737.76 (1), 787.94 (1), 834.99 (1), 877.44 (1), 922.60 (1), 966.34 (1), 1014.60 (1), 1051.30 (1),     & \citet{Hermes17}      \\
                    & 679.49 (2), 748.9 (2), 773.12 (2), 799.896 (2), 820.20 (2), 846.907 (2), 861.40 (2), 903.10 (2),      &                       \\
                    & 947.10 (2), 1095.434 (2)                                                                              &                       \\
                    & not used, central members of triplets: 368.624 ($\ell=1$,$m=0$), 461 ($\ell=1$,$m=0$)                               &                       \\
                    & not used, other: 505.447 (1), 557.244 (1), 404.603 (2)                                                &                       \\
\hline
\end{tabular}
 }
 \end{center}
\vspace{1mm}
\end{table}

\section{Spectroscopic data}
\label{sec:appendixB}

For reference, we list the spectroscopic measurements we have for the stars considered in this work, with sources cited. Most of that data was compiled with the help of the Montreal White Dwarf Database \citep{Dufour17}. We split the data into two tables for readability.

\begin{table}
  \begin{center}
  \caption{Mass and effective temperature data for the objects considered in this study, part 1.
     \label{tab:spectroscopy1}
}
 {\scriptsize
  \begin{tabular}{|lccccc|}
  \hline 
    Reference           & EC 14012-1446         & G38-29                & R808                  & GD 1212               & G29-38                \\
    \hline
    \hline
    \citet{Kilic20}     &                       & 10934 $\pm$ 144~K     & 11148 $\pm$ 43~K      &                       &                       \\
                        &                       & 0.521 $\pm$ 0.01 \msun & 0.598 $\pm$ 0.004 \msun &                    &                       \\
    \hline
    \citet{McCleery20}  &                       &                       & 11083.8 $\pm$ 88~K    &                       & 11295.9 $\pm$ 198~K   \\
                        &                       &                       & 0.565 $\pm$ 0.005 \msun &                     &0.595 $\pm$ 0.015      \\
    \hline
    \citet{Bedard17}    &                       &                       & 11187 $\pm$ 165~K     & 11037 $\pm$ 162~K     & 11315 $\pm$ 180~K     \\
                        &                       &                       & 0.61 $\pm$ 0.1 \msun  & 0.60 $\pm$ 0.03 \msun & 0.62 $\pm$ 0.08 \msun \\
    \hline
    \citet{Subasavage17} &                      &                       &                       &                       & 11240 $\pm$ 360~K     \\
                        &                       &                       &                       &                       & 0.6 $\pm$ 0.03 \msun  \\
    \hline
    \citet{Oswalt16}    &                       &                       &                       & 10678 $\pm$ 317~K     & 11956 $\pm$ 187~K     \\
                        &                       &                       &                       & 0.64 $\pm$ 0.02 \msun & 0.61 $\pm$ 0.03 \msun \\
    \hline
    \citet{Limoges15}   &                       &                       & 11240 $\pm$ 162~K     &                       & 12020 $\pm$ 83~K      \\
                        &                       &                       & 0.60 $\pm$ 0.03 \msun &                       & 0.69 $\pm$ 0.03 \msun \\
    \hline
    \citet{Hermes14a}    &                       &                       &                       & 10970 $\pm$ 165       & 10970 $\pm$ 165~K     \\
                        &                       &                       &                       & 0.6 $\pm$ 0.03 \msun  & 0.6 $\pm$ 0.03 \msun  \\
    \hline
    \citet{Xu14}        &                       &                       &                       &                       & 11820 $\pm$ 100~K     \\
                        &                       &                       &                       &                       & 0.85 \msun            \\
    \hline
    \citet{Giammichele12} &                     &                       &                       & 10938 $\pm$ 317~K     & 12206 $\pm$ 187~K     \\
                        &                       &                       &                       & 0.76 $\pm$ 0.02 \msun & 0.63 $\pm$ 0.03 \msun \\
    \hline
    \citet{Provencal12} & 12020~K               &                       &                       &                       &                       \\
                        & 0.69 \msun            &                       &                       &                       &                       \\
    \hline
    \citet{Gianninas11} & 12300 $\pm$ 190~K     & 11480 $\pm$ 174       & 11440 $\pm$ 168~K     &  11270 $\pm$ 165~K    & 12200 $\pm$ 187~K     \\
                        & 0.75 $\pm$ 0.03 \msun & 0.60 $\pm$ 0.03 \msun & 0.67 $\pm$ 0.03 \msun &  0.71 $\pm$ 0.03 \msun & 0.74 $\pm$ 0.03 \msun \\
    \hline
    \citet{Castanheira09} &                     &                       &                       &                       & 11910~K               \\
                        &                       &                       &                       &                       & 0.69 $\pm$ 0.03 \msun \\
    \hline
    \citet{Koester09}   &                       &                       &                       &                       & 11585 $\pm$ 8~K       \\
                        &                       &                       &                       &                       & 0.63 $\pm$ 0 \msun    \\
    \hline
    \citet{Subasavage09} &                      &                       &                       & 11000 $\pm$ 300~K     &                       \\
                        &                       &                       &                       & 0.76 $\pm$ 0.02       &                       \\
    \hline
    \citet{Gianninas06} &                       &                       &                       & 10970~K               &                       \\
                        &                       &                       &                       & 0.6 $\pm$ 0.03 \msun  &                       \\
    \hline
    \citet{Kilic06}     &                       &                       &                       &                       & 11600~K               \\
                        &                       &                       &                       &                       & 0.645 $\pm$ 0.035 \msun \\
    \hline
    \citet{Koester05}   &                       &                       &                       &                       & 12100~K               \\
                        &                       &                       &                       &                       & 0.53 $\pm$ 0.03 \msun \\
    \hline
    \citet{Liebert05}   &                       &                       & 11160 $\pm$ 161~K     &                       & 11820~K               \\
                        &                       &                       & 0.63 $\pm$ 0.03 \msun &                       & 0.70 $\pm$ 0.03 \msun \\
    \hline
    \citet{Zuckerman03} &                       &                       &                       &                       & 11600~K                 \\
                        &                       &                       &                       &                       & 0.615 $\pm$ 0.025 \msun \\
    \hline
    \citet{Koester01}   & 11458 $\pm$ 26~K      &                       &                       &                       & 11515~K $\pm$ 22~K     \\
                        & 0.635 $\pm$ 0.05 \msun &                      &                       &                       & 0.59 $\pm$ 0.005 \msun \\
    \hline
    \citet{McGraw76}    &                       &                       & 11120~K               &                       &                        \\
                        &                       &                       & 0.575 $\pm$ 0.025     &                       &                        \\
    \hline
    \citet{McGraw75}    &                       & 11160~K               &                       &                       & 11910~K                \\
                        &                       & 0.525 $\pm$ 0.025 \msun &                     &                       & 0.69 $\pm$ 0.03 \msun  \\
    \hline   
 \end{tabular}
 }
 \end{center}
\vspace{1mm}
\end{table}

\begin{table}
  \begin{center}
  \caption{Mass and effective temperature data for the objects considered in this study, part 2.
     \label{tab:spectroscopy2}
}
 {\scriptsize
  \begin{tabular}{|lccc|}
  \hline 
    Reference           & KIC 4552982              & KIC 210397465         & KIC 220453225      \\
    \hline
    \hline
    \citet{Hermes17}    &                       & 11520 $\pm$ 160~K     & 11540 $\pm$ 160       \\
                        &                       & 0.45 $\pm$ 0.04 \msun & 0.62 $\pm$ 0.04       \\ 
    \hline
    \citet{Bell15}      &   10860~K             &                       &                       \\
                        & 0.69 $\pm$ 0.04 \msun &                       &                       \\
    \hline
 \end{tabular}
 }
 \end{center}
\vspace{1mm}
\end{table}

\section{Period spacing diagrams}
\label{sec:appendixC}

In the main part of the paper, we showed the results of the statistical tests used to find any regular spacing in the period spectra considered in this work (Figs \ref{fig:01_dp_kic} and \ref{fig:02_dp_g29}). Here we include those for the rest of the object, for reference.

\begin{figure}[h!]
\epsscale{0.6}
\plotone{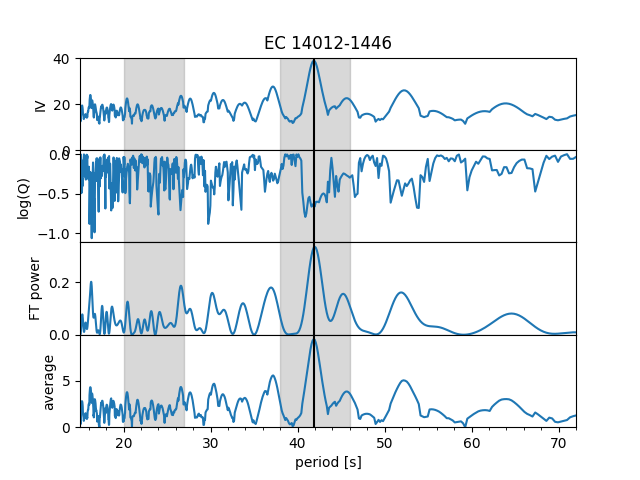}
\caption{Results of 3 statistical tests used in finding regular spacings in the period spectrum of EC 14012-1446. The 4th panel is an average of the above curves (IV, -log(Q), and FT power). The grey regions mark the period ranges where we would expect a regular spacing corresponding to the asymptotic limit for $\ell=1$ modes at the higher period range and for $\ell=2$ modes at the lower period range.
\label{fig:A1_dp_ec14012}
}
\end{figure}

\begin{figure}[h!]
\epsscale{0.6}
\plotone{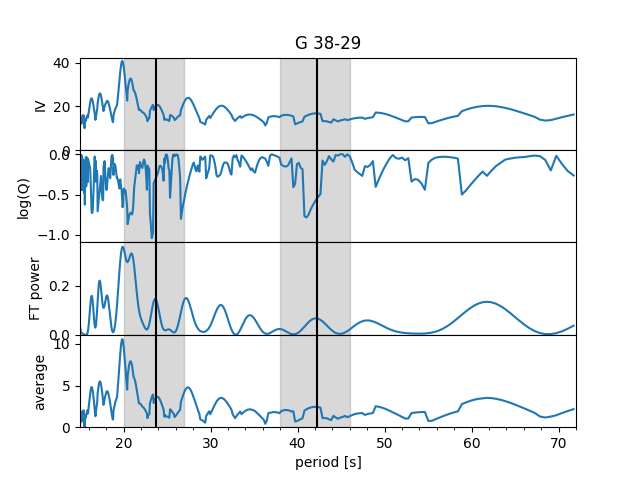}
\caption{The same as Fig. \ref{fig:A1_dp_ec14012} for G 38-29. We do not have a strong signal in the $\ell=1$ region.
\label{fig:A2_dp_g38}
}
\end{figure}

\begin{figure}[h!]
\epsscale{0.6}
\plotone{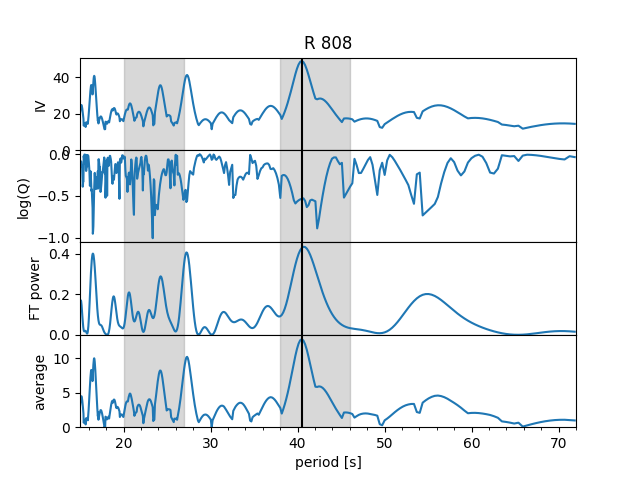}
\caption{The same as Fig. \ref{fig:A1_dp_ec14012} for R 808 ($\ell=1$ modes). The $\ell=1$ sequence is anchored by 4 triplets.
\label{fig:A3_dp_r808_ell1}
}
\end{figure}

\begin{figure}[h!]
\epsscale{0.6}
\plotone{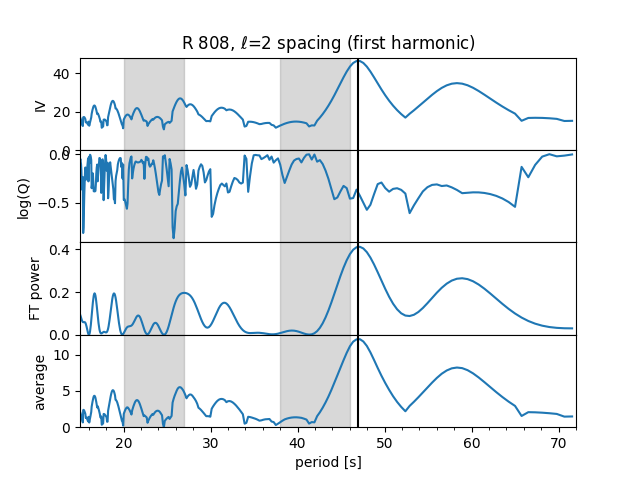}
\caption{The same as Fig. \ref{fig:A3_dp_r808_ell1} for the modes not belonging to the identified $\ell=1$ sequence. The tall peak at $\sim$ 47~s can be interpreted as twice an $\ell=2$ period spacing.
\label{fig:A4_dp_r808_ell2}
}
\end{figure}

\begin{figure}[h!]
\epsscale{0.6}
\plotone{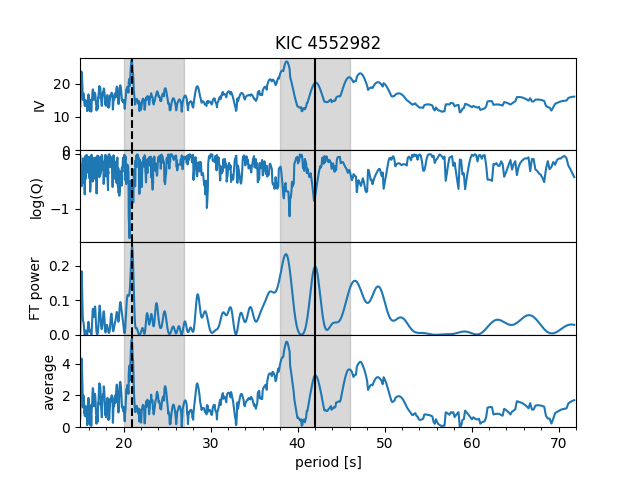}
\caption{The same as Fig. \ref{fig:A1_dp_ec14012} for KIC 4552982. We recover the $\sim$ 42~s spacing found by \citet{Bell15}. We interpret the (noisy) peak in the $\ell=2$ spacing region as a harmonic of the $\ell=1$ spacing (half of 42~s).
\label{fig:A5_dp_kic455}
}
\end{figure}

\begin{figure}[h!]
\epsscale{0.6}
\plotone{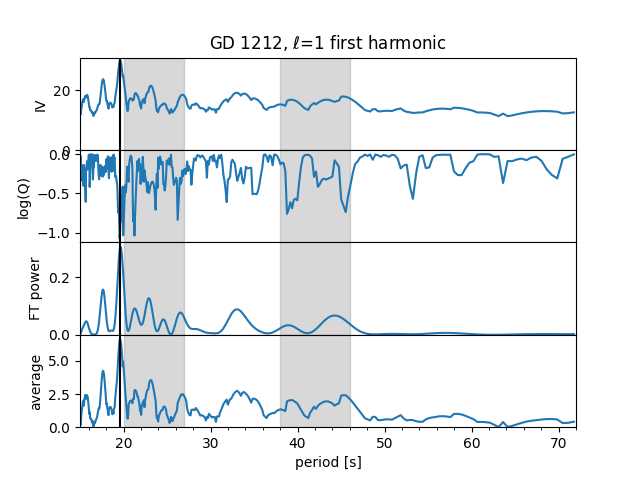}
\caption{The same as Fig. \ref{fig:A1_dp_ec14012} for GD 1212. The only significant peak, just below 20~s, may be a harmonic of the 42~s spacing found in the literature \citet{Hermes14a}. Otherwise, we fail to detect any regular pattern in this period spectrum.
\label{fig:A6_dp_gd1212}
}
\end{figure}

\begin{figure}[h!]
\epsscale{0.6}
\plotone{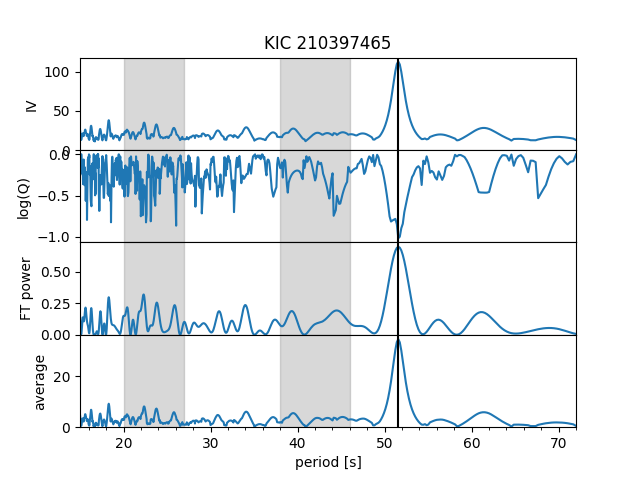}
\caption{The same as Fig. \ref{fig:A1_dp_ec14012} for KIC 210397565. For this object, the $\ell=1$ spacing is outside the expected range. This points to a below average mass, corroborated by the spectroscopy.
\label{fig:A7_dp_kic210}
}
\end{figure}

\begin{figure}[h!]
\epsscale{0.6}
\plotone{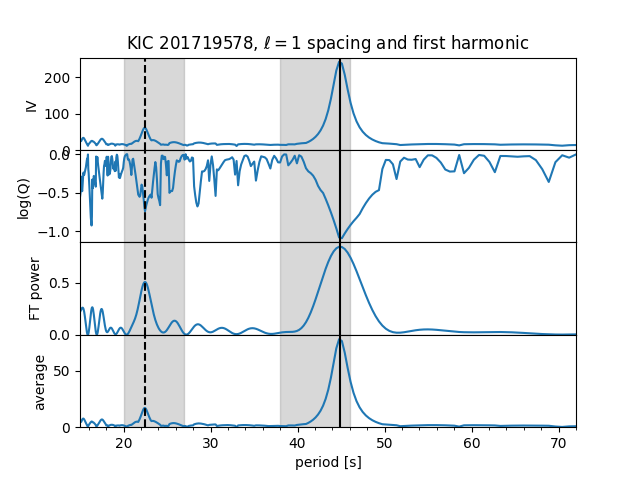}
\caption{The same as Fig. \ref{fig:A1_dp_ec14012} for KIC 201719578. This object has a rich period spectrum with well identified $\ell=1$ and $\ell=2$ sequences. We graph here the result of the statistical test for the higher $k$ modes in the $\ell=1$ sequence. The $\ell=1$ period spacing (45~s) and its harmonic produce strong signals.
\label{fig:A8_dp_kic2017_ell1}
}
\end{figure}

\begin{figure}[h!]
\epsscale{0.6}
\plotone{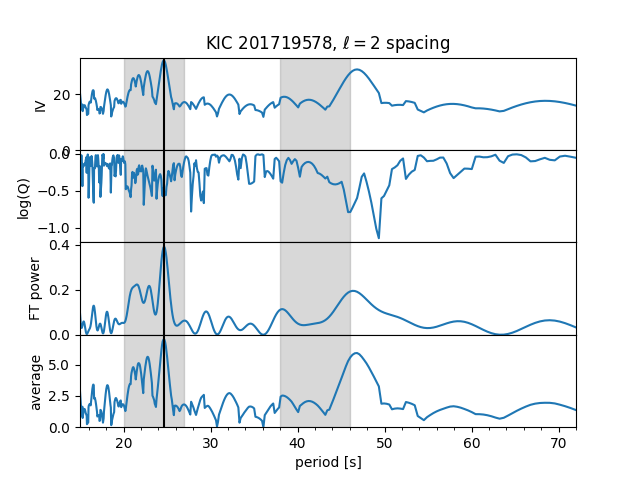}
\caption{The same as Fig. \ref{fig:A8_dp_kic2017_ell1} for the $\ell=2$ sequence. 
\label{fig:A9_dp_kic2017_ell2}
}
\end{figure}

\section{Mass-radius relations}
\label{sec:appendixD}

We provide here the mass-radius relations we used in this work. They are applicable to DA white dwarfs, with temperatures ranging between 9,000 and 50,000~K, and masses ranging between 0.45 and 1.0 \msun. A polynomial fit to WDEC models yields the following mass-radius relationship:

\begin{equation}
\label{eq:massradiusrel}
    \frac{M_{WD}}{M_\odot}=a(T) (\log R)^3+b(T) (\log R)^2 + c(T) \log R + d(T) ,
\end{equation}

\noindent where the radius R is in centimeters. The parameters a, b, c, and d are themselves cubic functions of effective temperature. We provide the coefficients of those cubic functions in table \ref{tab:massradiuspars}. As discussed in section \ref{sec:mass-radius}, we find a significant dependence of the mass-radius relation on the envelope mass \menv of the models and so we provide mass-radius relations for different values of \menv. We also specify the hydrogen layer thickness \mh chosen in each case. The rest of the parameters were set to reproduce the chemical profiles from \citet{Althaus10}. They change with the mass, but they also have a negligible effect on the radii of the models and so do not affect the mass-radius relation. The parameterized mass-radius relationships presented here reproduce the log or the radius (in cm) for any given mass white dwarf to better than 1\%, and in most cases to a fraction of a percent.

\begin{table}
  \begin{center}
  \caption{Parameters for the mass-radius relation (Eqn. \ref{eq:massradiusrel} and subsequent text). They are given for different envelope masses. The parameters are valid for $9,000~K<\rm{T_{eff}}<50,000~K$ and $0.45<\rm{M_*/M_\odot}<1.0$, with the exception of the set for \menv = $10^{-1.8}$, valid for masses up to 0.7 \msunns.
  \label{tab:massradiuspars}
}
 {\scriptsize
  \begin{tabular}{ccccc}
    Coefficient & $T^3$ & $T^2$ & $T^1$ & $T^0$                                     \\
    \hline 
    \multicolumn{5}{c}{\menv = $10^{-1.8}$, \mh = $10^{-4.0}$}                      \\
    a &   4.10323909e-13  & -3.46676092e-08 &  6.19733751e-04  &  4.92048413e+00    \\
    b &  -1.10282233e-11  &  9.30898701e-07 & -1.65965255e-02  & -1.31535950e+02    \\
    c &   9.87956049e-11  & -8.33183110e-06 &  1.48161212e-01  &  1.16988911e+03    \\
    d &  -2.95000724e-10  &  2.48563818e-05 & -4.40913813e-01  & -3.46117336e+03    \\
    \hline 
    \multicolumn{5}{c}{\menv = $10^{-2.2}$, \mh = $10^{-4.5}$}                      \\
    a &  -5.57603801e-14  &  4.13014428e-09 &  -2.06293524e-04 &  7.90051492e+00    \\
    b &   1.51369845e-12  & -1.11932714e-07 &   5.55243171e-03 & -2.11549912e+02    \\
    c &  -1.36923548e-11  &  1.01066588e-06 &  -4.97819191e-02 &  1.88591242e+03    \\
    d &   4.12715285e-11  & -3.04040695e-06 &   1.48686784e-01 & -5.59664051e+03    \\
    \hline 
    \multicolumn{5}{c}{\menv = $10^{-3.0}$, \mh = $10^{-4.5}$}                      \\
    a &   6.80952161e-15  & -1.06529508e-09 & -7.69095813e-05 &  8.77613909e+00     \\
    b &  -1.45115599e-13  & -2.46315438e-08 &  2.18461489e-03 & -2.35343927e+02     \\
    c &   9.64368024e-13  & -1.85355135e-07 & -2.05906086e-02 &  2.10133864e+03     \\
    d &  -1.88927110e-12  &  4.49638552e-07 &  6.44336645e-02 & -6.24656119e+03     \\
   \hline
   \multicolumn{5}{c}{\menv = $10^{-5.0}$, \mh = $10^{-5.0}$}                       \\
   a & -1.88964308e-13  &  1.69031370e-08   & -5.37836472e-04 &  1.19395906e+01     \\
   b &  5.05614114e-12  & -4.52625367e-07   &  1.44180326e-02 & -3.19164038e+02     \\
   c & -4.50918721e-11  &  4.03960633e-06   & -1.28807062e-01 &  2.84153754e+03     \\
   d &  1.34034729e-10  & -1.20162358e-05   &  3.83491502e-01 & -8.42509531e+03     \\
   \hline
   \multicolumn{5}{c}{\menv = $10^{-8.5}$, \mh = $10^{-10.0}$}                      \\
   a & -3.96541747e-14  &  6.82299143e-10   & -8.76654844e-06 &  7.55260796e+00     \\
   b &  1.06694927e-12  & -1.94508661e-08   &  2.99584319e-04 & -2.02098726e+02     \\
   c & -9.56791571e-12  &  1.83980663e-07   & -3.23168131e-03 &  1.80030626e+03     \\
   d &  2.85964199e-11  & -5.77728986e-07   &  1.12145010e-02 & -5.33819929e+03     \\
   \hline
  \end{tabular}
  }
 \end{center}
\vspace{1mm}
\end{table}

\end{appendix}



\end{document}